\documentclass[prl,twocolumn,preprintnumbers,amsmath,amssymb, nofootinbib]{revtex4-1}
\usepackage{color}
\usepackage[active]{srcltx}
\usepackage{amsmath,amsfonts,amssymb,amsthm,amstext,amscd,eucal,srcltx}
\usepackage{epsfig,graphicx,bm}
\usepackage{epstopdf, epsf}
\usepackage{dcolumn}% Align table columns on decimal point
\usepackage{hyperref}

\def\nn{\nonumber}

\newcommand{\mpl}{M_{\rm pl}}
\def\AS{{\alpha_k^S}}
\def\A{{\alpha_k}}
 \def\B{{\beta_k}}
 \def\AT{{\alpha_k^T}}
 \def\BS{{\beta_k^S}}
\def\BT{{\beta_k^T}}
\def\chit{\chi_{{}_T}}
\def\chis{\chi_{{}_S}}
\def\phit{\varphi_{{}_T}}
\def\phis{\varphi_{{}_S}}

\newcommand{\be}{\begin{equation}}
\newcommand{\ee}{\end{equation}}
\newcommand{\beqa}{\begin{eqnarray}}
\newcommand{\eeqa}{\end{eqnarray}}
\newcommand{\bi}{\begin{itemize}}
\newcommand{\ei}{\end{itemize}}

\begin{document}
\title{Non-Bunch-Davies Initial State Reconciles Chaotic Models with BICEP and Planck}
\author{Amjad Ashoorioon$^1$, Konstantinos Dimopoulos$^1$, Mohammad M. Sheikh-Jabbari$^{2,3}$ and Gary Shiu$^{4,5}$}
\bigskip\medskip
\affiliation{$^1$~Consortium for Fundamental Physics, Physics Department, Lancaster University, LA1 4YB, United Kingdom. \\
$^2$ School of Physics, Institute for Research in Fundamental Sciences (IPM), P .O. Box 19395-5531, Tehran, Iran and\\
$^3$ Department of Physics, Kyung Hee University, Seoul 130-071, Republic of Korea\\
$^4$ Department of Physics, University of Wisconsin, Madison, WI 53706, United States\\
$^5$ Center for Fundamental Physics and Institute for Advanced Study,
Hong Kong University of Science and Technology, Hong Kong }
\vfil

\begin{abstract}{The BICEP2 experiment has announced a signal for primordial gravity waves with tensor-to-scalar ratio $r=0.2^{+0.07}_{-0.05}$ \cite{BICEP}. There are two ways  to reconcile this result with the latest Planck experiment \cite{Planck-data}. One is by assuming that there is a considerable tilt of $r$, $\mathcal{T}_r$, with a positive sign,  $\mathcal{T}_r=d\ln r/d\ln k\gtrsim 0.57^{+0.29}_{-0.27}$ corresponding to a blue tilt for the tensor modes of order $n_T\simeq0.53 ^{+0.29}_{-0.27}$,  assuming the Planck experiment best-fit value for  tilt of scalar power spectrum $n_S$. The other possibility is to assume that there is a negative running in the scalar spectral index, $dn_S/d\ln k\simeq -0.02$ which pushes up the upper bound on $r$ from $0.11$  up to $0.26$ in the Planck analysis assuming the existence of a tensor spectrum. Simple slow-roll models fail to provide such large values for $\mathcal{T}_r$ or negative runnings in $n_S$ \cite{BICEP}. In this note we show that a non-Bunch-Davies initial state for perturbations can provide a match between large field chaotic models (like $m^2\phi^2$) with the latest Planck result \cite{ADSS} and BICEP2 results by accommodating either the blue tilt of $r$ or the negative large running of $n_S$.
}
\end{abstract}


\preprint{\today }

\maketitle

%\section{Introduction}
%\fixme{Introduction should be modified....}

Early Universe cosmology has become a very active area of research
in the last decade or so, as there is a wealth of precise cosmic microwave background (CMB)  measurements  pouring in. In particular, since last year
two major Collaborations Planck \cite{Planck-data} and BICEP \cite{BICEP} have announced their results. The CMB measurements analyzed with other cosmological data favor the simple $\Lambda$CDM model for late time cosmology and inflationary paradigm for early stages of Universe evolution.
According to the Planck Collaboration data \cite{Planck-data} the power spectrum of CMB temperature fluctuations (or as it is known, the power spectrum of curvature perturbations) ${\cal P}_S$ is measured to be about $2.195\times 10^{-9}$. The spectrum is almost flat, with a few-percent tilt toward larger scales (i.e., red spectrum) and is almost Gaussian.

Planck took cosmologists by surprise as it not only did not observe non-Gaussianity, which could have been used to considerably constrain inflationary models, but also put a strong upper bound on the amplitude of primordial gravity waves during inflation. These gravity waves are tensor mode fluctuations which are produced during inflation. The power spectrum of gravity waves ${\cal P}_T$ is usually reported through the tensor-to-scalar ratio $r={\cal P}_T/{\cal P}_S$ which Planck reported to be bounded at $2\sigma$ level as $r<0.12$, assuming no running in the scalar spectral index, $n_S$. This bound corresponds to the pivot scale $k_{\ast}=0.002Mpc^{-1}$, $\ell_{\ast}\simeq 28$. The tilt in the power spectrum of curvature perturbations is customarily denoted by $n_S-1$, $n_S-1\equiv d{\ln P}_S/d{\ln k}$, where $k$ is inverse of the scale. Planck constrained $n_S-1=-0.0397\pm 0.0146$ at 2$\sigma$ level. Of course the upper bound on $r_{0.002}$ could be increased, if there is running in the scalar spectral index. The Planck Collaboration limits the running of scalar spectral index, $d n_S/d \ln k=-0.021\pm 0.011$ in the presence of the tensor modes. Then the upper bound on $r$ at the Planck pivot scale, $k_{\ast}=0.002Mpc^{-1}$, becomes weaker, $r<0.26$. Planck's measurement of $n_S$ and its running $dn_S/d\ln k$  disfavored many single field models, especially those with convex potential \cite{Planck-data}.

CMB besides having one-in-$10^5$ part temperature fluctuations is partially polarized and the parity odd polarization, the B-mode, is usually attributed to primordial gravity waves, tensor modes \cite{Starobinsky:1979ty}. BICEP2 Collaboration has recently announced observation of B-mode polarization \cite{BICEP}. BICEP results took cosmologists by an even greater surprise, when measured $r=0.2^{+0.07}_{-0.05}$\footnote{The BICEP experiment has interpreted the signal as solely given rise from the primordial B-mode during inflation, underestimating the contribution of combination of Galactic foregrounds and lensed E-modes. This assumption was recently questioned in \cite{Flauger:2014qra} where the authors showed that the data is consistent with both $r=0.2$ and negligible foregrounds and also with $r=0$ and a significant dust polarization signal. The analysis of this paper is relevant if the value of $r$ at $\ell=80$ obtained after ``realistic'' foreground removal turns out to be larger than the corresponding value of $r$ at the Planck experiment pivot scale.}. This was not an outright inconsistency between the two collaborations though, because BICEP focused on smaller scales than the range of scales covered by Planck; BICEP data is for $\ell\sim 80$. BICEP result was challenging in view of Planck results, as the measured value is already in the region which was excluded by Planck, unless either ({\it i}) the power spectrum of gravity waves considerably grows as we move to smaller scales, {\it i.e.} a blue, with relatively large tilt, for power spectrum of tensor modes, or ({\it ii}) there is a large negative running in the scalar spectral index. These are two possibilities to reconcile BICEP data with Planck results \cite{BICEP}. Nonetheless, both potential ways for Planck-BICEP reconciliation seem very hard to achieve in the context of slow-roll inflationary models composed of scalar fields minimally coupled to Einstein gravity. To see the difficulties associated with these options, we need to go through the equations more closely.

In the first approach, the controversy is best formulated in terms of  the tilt of tensor-to-scalar ratio ${\cal T}_r$,
\be\label{Tr-spectral-tilts}
\mathcal{T}_r\equiv \frac{d\ln r}{d\ln k}=\frac{d\ln \mathcal{P}_T}{d\ln k}-\frac{d\ln \mathcal{P}_S}{d\ln k}=n_T-(n_S-1)\,,
\ee
where $n_T$ is the tilt of power spectrum of tensor modes and $n_S-1$ is the tilt of the power spectrum of curvature perturbations. Planck requires $n_S-1$ to be negative and of order $-0.04$. Standard, textbook analysis for slow-roll inflationary models leads to  the ``consistency relation'' $n_T=-r/8$ \cite{Starobinsky:1985ww}, which is a red-tilt for gravity waves \cite{Mukhanov,Lyth:2009zz}. Therefore, $n_T$, too, is negative and of order $ \mathcal{O}(-0.01)$ for such inflationary models. On the other hand, BICEP-Planck reconciliation requires
\be
{\cal T}_r \geq +0.30\,.
\ee
The lower bound of the inequality corresponds to the lower end of the $1\sigma$ interval of the BICEP results, $r=0.15$,  which is already smaller than the more conservative tensor to scalar ratio, r=0.16, quoted in BICEP, obtained from the best data driven model of the emission of polarized dust. This clearly shows the tension between standard slow-roll models, and in particular the consistency relation with Planck+BICEP data: Slow-roll inflationary models cannot easily and readily accommodate the respectively large value of tensor-to-scalar spectral tilt ${\cal T}_r$ and the blue tensor spectrum required by recent observations (please see \cite{Gong:2007ha} for another attempt to make $r$ run).

As stated above, another way to conciliate these two experiments is by assuming a running spectral index. However in the presence of Bunch-Davies initial states, such a running in slow-roll models is second order in terms of slow-roll parameters \cite{Planck-data}
\be\label{running-of-tilt}
\frac{d n_S}{d\ln k}=16\epsilon \eta-24 \epsilon^2-2\xi^2,
\ee
where $\epsilon$, $\eta$ are the usual first and second slow-roll parameters defined in \eqref{slow-roll-def} and $\xi$ is the third slow-roll parameter defined as
%\cite{Planck-data}
\be
\xi^2\equiv \frac{M_{\rm Pl}^4 V_{\phi}V_{\phi\phi\phi}}{V^2}.
\ee
The running of scalar spectral index can be also achieved by assuming a scale and space-dependent modulation which suppresses the CMB power spectrum at low multipoles \cite{McDonald:2014kia}.

 The possibility which we will entertain here to achieve either of these goals is based on the fact that in deriving standard cosmic perturbation theory results, besides the action of the model (which establishes the background inflationary dynamics and provides the equation of motion for cosmic perturbation fields), we also need to specify the initial quantum state over which these (quantum) cosmic perturbations have been produced. The standard initial state used is the Bunch-Davies (BD) vacuum state \cite{Bunch:1978yq}, stating that perturbation modes with physical momenta much larger than the Hubble scale during inflation $H$, effectively propagate in a vacuum state associated with flat space, the standard quantum field theory vacuum state.

In the context of first approach, in particular noting \eqref{Tr-spectral-tilts}, to remedy Planck-BICEP tension we need to relax the consistency relation $n_T=-r/8$. Considering non-Bunch-Davies (non-BD) initial state for cosmic perturbations during inflation provides the setup to relax the consistency relation \cite{Hui:2001ce} (see \cite{Initial-data-literature} for some earlier works on the non-BD inflationary cosmology.) In fact, in our previous paper \cite{ADSS} we discussed such a setup and already used it in resolving the tension between Planck data and large-field chaotic inflationary models, including the simplest inflationary model with $m^2\phi^2$ potential for the inflaton field $\phi$. Large-field models generically predict large value for tensor-to-scalar ration $r$, with $r\sim 0.05-0.2$ \cite{Linde:1983gd}. So, they are potentially very good candidates for accommodating BICEP too. As we will discuss here, non-BD initial state can equip the large-field models with the tilt of $r$, ${\cal T}_r$, (equivalent with blue tensor spectrum, $n_T>0$) or the negative running of $dn_S/d\ln k\simeq -{\mathrm few}\times 0.01$ needed for BICEP-Planck reconciliation; the chaotic model $m^2\phi^2$ \cite{Linde:1983gd} with non-BD initial state nicely fits with all available cosmological data.

The rest of this Letter is organized as follows. We first briefly review the setup presented in \cite{ADSS} to fix our notations. We then show that a mild tilt in the non-BD initial state will accommodate BICEP as well as Planck data. We first focus on the possibility of producing a running $r$ and then try to resolve these two experiments conflict with negative large running $n_S$. In the end we make some concluding remarks.

\paragraph{Power spectra and non-BD initial state.}
Here we consider a simple single-field slow-roll inflationary model described by the action
\begin{equation}
\mathcal{L}=-\frac{\mpl^2}{2}R-\frac{1}{2}\partial _{\mu }\phi \partial ^{\mu }\phi -V(\phi)\,,
\label{action-scalar-minimal}
\end{equation}
where $\mpl=(8\pi G_N)^{-1/2}=2.43\times 10^{18}$ GeV is the reduced Planck mass. We take our model to be a chaotic inflation large-field model \cite{Mukhanov}, motivated by the recent observation of tensor modes \cite{BICEP}, e.g. $V(\phi)=\frac12m^2\phi^2$. The details of cosmic perturbation theory analysis for this model in standard Bunch-Davies vacuum may be found in standard textbooks, e.g. \cite{Mukhanov}, and the modifications due to non-BD initial state in \cite{Initial-data-literature,ADSS}. For completeness we have gathered a summary of this analysis in the appendix. The power spectra and tensor-to-scalar ratio, $r$, are
\be\label{power-spectra&r}
\begin{split}
{\cal P}_S &=\frac{1}{8\pi^2\epsilon}\left(\frac{H}{\mpl}\right)^2\ \gamma_S\cr
{\mathcal P}_T &=\frac{2}{\pi^2}\left(\frac{H}{\mpl}\right)^2\,\ \gamma_T\cr
r&=\frac{{\cal P}_T}{{\cal P}_S}=16\epsilon\ \gamma\,,\end{split}
\ee
with
\be\label{gammaS-gammaT}
\gamma_S=|\AS-\BS|^2_{{}_{k={\cal H}}},\quad \gamma_T=|\AT-\BT|^2_{{}_{k={\cal H}}},\quad \gamma=\frac{\gamma_T}{\gamma_S},
\ee
where $\alpha$'s and $\beta$'s parameterize non-BD initial state for scalar and tensor modes and
the spectral tilts are then\footnote{It is instructive to note and recall expressions for the tilts of power spectra and scalar-to-tensor ratio $r$ for $\lambda\phi^n$ chaotic models \emph{in the BD vacuum}. For these models $\eta=2(n-1)\epsilon/n$, and
\be
(\mathcal{T}_r)_{\text{BD}}=+\frac{4}{n}\epsilon\,,\qquad (n_S-1)_{\text{BD}}=-\frac{2(n+2)}{n}\epsilon\,.
\ee
Noting that $r\propto \epsilon\propto (n_S-1)$, one can relate the tilt of $r$ to the running of the spectral tilt \eqref{running-of-tilt}. Explicitly,
\be
(\mathcal{T}_r)_{\text{BD}}=\frac{\ln (1-n_S)}{d\ln k}=\frac{1}{n_S-1}\frac{d n_S}{d\ln k}\,.
\ee
}
\be\label{tilts}
\begin{split}
n_S-1&= (n_S-1)_{\text{BD}}+\frac{d\ln\gamma_S}{d\ln k}\,,\cr
n_T&= (n_T)_{\text{BD}}+\frac{d\ln\gamma_T}{d\ln k}\,,\cr
{\cal T}_r&= (\mathcal{T}_r)_{\text{BD}}+\frac{d\ln\gamma_T}{d\ln k}-\frac{d\ln\gamma_S}{d\ln k}\,.
\end{split}
\ee

The Lyth bound \cite{Lyth} and the consistency relation will also be modified due to the non-BD effects to \cite{ADSS}
\be\label{modified-Lyth}
r\lesssim 2.5\times 10^{-3} \left(\frac{\Delta\phi}{\mpl}\right)^2\ {\gamma}\,,\qquad r=-8n_T \gamma\,,
\ee
where $\Delta\phi$ is the inflaton field displacement during inflation. The modification in the consistency relation is, as discussed, what can resolve the mismatch of slow-roll models with BICEP+Planck data.\footnote{As we discussed in \cite{ADSS}, in major part of the constrained non-BD parameter space, $\gamma\leq 1$ and effective field theory could not be saved by reducing $\Delta \phi<\mpl$, enhancing $\gamma$.}

\paragraph{Parameterizing the initial states.}  We  note the fact that
only the phase difference between the Bogoliubov coefficients $\alpha_k$ and $\beta_k$ appears in the power spectra and their $k$-dependence (\emph{cf.}\eqref{gammaS-gammaT}). Moreover, the normalization conditions \eqref{Wronskian} and \eqref{normalization-Tensor}, too, depend only on the phase difference. Therefore, one can take out the average (an overall phase) and parameterize the coefficients such that only the phase difference appears \cite{ADSS}:
%\footnote{\textbf{Please note that only the phase difference between $\alpha$ and $\beta$ enter into the physical quantities. Here we have parameterized our %Bogoliubov coefficients such that the phases are minus each other. This way the phase difference between the two coefficients is $2\varphi$, which can be a %scale-dependent quantity. This way we will simplify some of the equations considerably.}}
\be\label{parametrization}\begin{split}
\AS =\cosh\chis e^{i\phis}\,,\qquad \BS =\sinh\chis e^{-i\phis}\,\cr
\AT =\cosh\chit e^{i\phit}\,,\qquad \BT =\sinh\chit e^{-i\phit}\,.
\end{split}
\ee
We consider  a crude model in which \cite{Boyanovsky:2006qi},
\be\label{beta-Gaussian}
|\beta_k^{\{S,T\}}|\propto \beta_0^{\{S,T\}} \exp\left\{{{-k^2/\left[M a(\tau)\right]}^{2}}\right\}
\ee
(or any smooth function in which $\left|\beta_k\right|^2$ falls off as $k^{-(4+\delta)}$). Here $M$ is a super-Hubble energy scale associated with the new physics which leads to the non-BD initial state. In this scenario, all the $k$ modes are pumped to an excited state as their physical momentum reaches the cutoff $\frac{k}{a(\tau)}=M$. The choice in \eqref{beta-Gaussian} indicates that $M$ is the (cutoff) scale at which the mode gets excited from Bunch-Davies vacuum.

The physically allowed region in the four parameter space of initial states is subject to the following constraints: (1) Absence of backreaction of initial states on the inflationary background; (2) Planck normalization for ${\cal P}_S$; (3) value of spectral tilt $n_S-1$ as observed by Planck; (4) fitting the value of $r$ and the corresponding tilt ${\cal T}_r$, as required by BICEP+Planck, {\it i.e.} we take $r_{Planck}\leq 0.12$ (at ${\ell}_{\ast}\simeq 28$) and $r_{BICEP}\simeq 0.2$ (at $\ell\sim 80$). In our analysis we focus on large-field single-field slow-roll models. The first three conditions were also considered in \cite{ADSS} while the fourth one is new.

{Absence of backreaction of initial excited state on the background slow-roll inflation trajectory} implies that the energy stored in the initial non-BD state for both scalar and tensor sectors should not exceed the change in the energy density in one e-fold.  This condition is fulfilled if \cite{ADSS}
\begin{eqnarray}\label{beta-scalar-backreaction}
   \sinh\chis \lesssim  \epsilon\frac{H M_{\rm Pl}}{M^2}\,,\quad \sinh\chit \lesssim  \epsilon\frac{H M_{\rm Pl}}{M^2}.
\end{eqnarray}
The above indicates that the upper bound on the deviation from BD initial state measured by $\chis$ is inversely proportional to the scale of new physics $M$. Hence, larger values of $M$ require smaller $\chis$. {The COBE normalization} implies
\be\label{H/Mpl}
\frac{H}{\mpl}=\frac{1}{\sqrt\gamma_S} 3.78\times 10^{-5}\,.
\ee

{Assuming $n_S$ takes its best fit value of Planck, $n_S-1\simeq-0.04$,} and that $\epsilon\sim 0.01$, then $d\ln \gamma_S/d\ln k\lesssim 10^{-2}$.

The above conditions are achieved if
we take $\chit$ and $\chis$ to take typical values \cite{ADSS}, i.e. $\sinh\chis\simeq e^{\chis}/2\,,\ \sinh\chit\simeq e^{\chit}/2$ and hence
$$
\gamma_S\simeq e^{2\chis} \sin^2\phis\,,\qquad \gamma_T\simeq e^{2\chit} \sin^2\phit\,.
$$
Moreover to be able to rely the effective field theory methods, we are typically interested in larger values of $M$ which is possible if $\phis$ is close to maximal; $M\simeq 20H$ happens when $\phis\sim\pi/2$ \cite{ADSS}.

\paragraph{Blue tensor spectrum.} In this approach, to reconcile BICEP+Planck we want $n_S-1\sim -0.04$ and $\mathcal{T}_r\geq +0.3$ and the Planck bound on $r$ requires $\gamma<3/4$.
Therefore,
\be
\begin{split}
e^{2(\chit-\chis)}\sin^2\phit<3/4,\cr
\frac{d\chis}{d\ln k}\lesssim 10^{-2},\cr
\frac{d\chit}{d\ln k}+\cot\phit\frac{d\phit}{d\ln k}\gtrsim 0.13.
\end{split}
\ee
We need not impose any condition on $\frac{d\phis}{d\ln k}$, as $\partial \ln \gamma_S/\partial \phis=0$ at $\phis=\frac{\pi}{2}$. Above we have also assumed that $\tan{\phit}\gg e^{-2\chit}$. If $\chit\gtrsim 1$, in principle very small values for $\phit$ could be achieved.

One theoretically interesting option is to have $\chit=\chis$,  corresponding to the case where the numbers of particles in the tensor and scalar excited states are equal. Change in $n_S-1$ from its Bunch-Davies value could be set to zero, if $d\chis/d\ln k=0$. This choice is particularly useful for $m^2\phi^2$ as its spectral index with Bunch-Davies vacuum nicely matches the Planck results. Since $\chis=\chit$, one has to assume that $\chit$ is scale independent too. A positive tensor spectral index would come totally from the scale-dependence of $\phit$. The amount of suppression of $r_{0.002}$, will be equal to $\sin^2\phit$, while at BICEP scales ($\ell\sim 80$) $\phit$ is close to its maximal value $\pi/2$. In such a scenario, to get $r=0.12$ at $\ell\sim 28$ we need
\be
\phit\sim \frac{\pi}{3}\simeq 1.04, \quad \text{at}\ \ell\simeq 28,
\ee
and the variation of $\phit$ with scale has to be
\be
\frac{d\phit}{d\ln k}\simeq 0.5.
\ee
{Thus $\phit$ has to be scale dependent such that $\phit\propto k^{0.5}$}. Asking for larger suppression of $r$ at $\ell\sim28$ would require smaller values of $\phit$, and hence larger values of  $\frac{d\phit}{d\ln k}$. {For example the case $\phit=-k\tau_0$, where $\tau_0$, is the preferred initial time, can provide larger suppression at $\ell\simeq 28$. Since larger values of logarithmic tilt of $\phit$ are not theoretically well motivated, getting small $\phit$ values for the $\chit=\chis$ scenario is not a feasible option.}

The other possibility to obtain positive $\mathcal{T}_r$ and hence a blue tensor spectrum is to allow for running of $\chit$. If this is the sheer cause of a blue gravitational spectrum a value of
 \be
 \frac{d\chit}{d\ln k}\gtrsim 0.13
 \ee
is required to solve the discrepancy between BICEP and Planck data. Depending on the value of $\phit$, one has to ensure the required suppression through the $\gamma$ factor.

 \paragraph{Running scalar spectral index.} One can produce such a negative large running with scale-dependent excited states too. In this case there is no need to suppress the prediction of a model like $m^2\phi^2$ for $r$ at the Planck pivot scale, {\it i.e.} we can assume that $\phit=\pi/2$. Only the running of scalar spectral index of order $-\mathrm{few}\times 0.01$ would be enough to patch up two experiments. The running of scalar spectral index in the presence of excited states is
\be
\frac{d n_S}{d \ln k}=\left(\frac{d \chi_S}{d \ln k}\right)_{BD}+\frac{d^2 \ln \gamma_S}{(d\ln k)^2}
\ee
 The contribution of the scale-dependent Bogoliubov coefficients to the running of scalar spectral index close to $\phis=\frac{\pi}{2}$ is
\be
\frac{d^2 \ln \gamma_S}{(d\ln k)^2}=2 \frac{d^2  \chi_S}{(d\ln k)^2}-2 {\left(\frac{d \phis}{d \ln k}\right)}^2.
\ee
Now there are two ways one can achieve the negative running of order $-0.02$:
\begin{itemize}
\item One can assume that $\frac{d^2  \chi_S}{(d\ln k)^2}\simeq -0.01$. Then the desired running in the scalar spectral index could be achieved. {This, for example, would correspond to the case where $|\BS|$ decreases slowly and quadratically with $\ln k$ and could be achieved  if $|\BS| \propto -0.005 (\ln k)^2$. The phase, $\phit$, can be constant in this case.}

\item Instead one can assume that $\left(\frac{d \phis}{d \ln k}\right)\simeq 0.1$. What is notable and interesting in this case is that the running always turns out to be negative. The required scale-dependent phase turns out to be quite small in this case too.  {The number density of the particles in the scalar perturbations could be scale-independent in this case.}
\end{itemize}

\section{Concluding remarks }

As discussed in \cite{ADSS}, non-Bunch-Davies initial condition for inflationary perturbations  with a typical value of the $\chis$ parameter ($\chis\gtrsim 1$) with the non-BD phase $\phis$ close to maximum,  $\phis\sim \pi/2$, can reconcile the $m^2\phi^2$ chaotic model with Planck data, $r_{0.002}<0.12$, if the scale of new physics which sources the non-BD initial state $M$, is around $20H$. Observation of B-modes by the BICEP experiment at $\ell\simeq 80$ can be matched with the bound from Planck data, either if the gravity wave spectrum has a blue tilt of order $0.53$ or there is a running of scalar spectral of order $d n_S/d \ln k\simeq -0.02$. Due to the large blue tilt for the gravity waves needed for this purpose, the second option is the preferred one.

Slow-roll inflation with BD initial condition cannot provide any of the above two possibilities. One can obtain such a blue spectrum for the gravity waves if  the tensor Bogoliubov coefficient has a maximum allowed value for the phase at the Planck pivot scale, $(\phit\simeq \pi/3)$, with moderate $k$-dependence, $\partial\phit/\partial \ln k\simeq 0.5$. The negative large running of scalar spectral index could be obtained if $d\phis/d\ln k\simeq 0.1$. In the second case, the running turns out to be always negative. Simple chaotic models, in particular $m^2\phi^2$ model, have been of interest because they are endowed with simplicity and beauty. As our analysis indicates they can be compatible with both Planck and BICEP results, if perturbations start in a non-BD initial state at the beginning of inflation.

In the current work we mainly focused on the non-BD initial state effects on observables related to two-point functions, the power spectra and their tilts. One should in principle also analyze the bi-spectra and non-Gaussianity in this context. Such an analysis has been carried out in many papers in the literature (see \cite{ADSS} and references therein). As we pointed out in \cite{ADSS}, such excited initial states can hardly leave observable signatures on non-gaussianity if the bound from backreaction is respected and the scale of new physics is separated maximally from the inflationary Hubble scale. The local configuration is the one which is mostly influenced in the presence of excited states for which the $f_{\rm NL}$ at most reaches $0.43$, which is well within the bounds allowed by the Planck experiment \cite{Planck-data}.

Noting that B-modes are coming from purely tensor perturbations of the metric \cite{Ashoorioon:2012kh} and that non-BD initial state for perturbations is provided from a high energy pre-inflationary physics, such resolutions may open up a window to the realm of quantum gravity, a territory which is untouchable by collider experiments. To that end, one should construct explicit models within the (existing) theoretical frameworks which can realize either of the two possibilities discussed here. We hope to return to this question in upcoming publications.

%\section*{Acknowledgements}

A.A. and K.D. (in part) are supported by the Lancaster-Manchester-Sheffield Consortium for Fundamental Physics under STFC grant ST/J000418/1.
GS is supported in part by DOE grant DE-FG-02-95ER40896.

\appendix
\section{Cosmological perturbations in non-BD  initial state}

To explore the effects of non-Bunch-Davies (non-BD) initial state of perturbations and setup our notations, we briefly review cosmic perturbation theory.
More detailed analysis in standard BD vacuum may be found in many textbooks e.g. \cite{Mukhanov}, more detailed discussion on non-BD may be found in \cite{ADSS} and references therein. Here we will consider slow-roll models described by the action \eqref{action-scalar-minimal}.

The space-time metric in presence of scalar and tensor perturbations can be parameterized as
\begin{equation}\label{mtrc}
ds^{2}=a^{2}(\tau )\left[ -(1+2\Phi ){d\tau }^{2}+\left( (1-2\Psi ){\delta }%
_{ij}+h_{ij}\right) d{y}^{i}d{y}^{j}\right] .\nn
\end{equation}%
$\Phi $ and $\Psi $ are the scalar Bardeen potentials which are equal for the scalar-driven inflationary model we are considering.
$h_{ij}$ is a symmetric
divergence-free traceless tensor field, $
h_{i}^{i}=0,\ \partial^ih_{ij}=0$.
The inflaton field also fluctuates around it homogeneous background value
\begin{equation}
\phi (\tau )=\phi_{\rm hom.}(\tau )+\delta \phi .  \label{inflat}
\end{equation}%
where
$\phi_{\rm hom.}(\tau)$ is the homogeneous part of the inflation which satisfies $\delta
\phi \ll \phi_{\rm hom.}(\tau)$. For the slow-roll a quasi-de-Sitter inflationary trajectories
\begin{eqnarray}
%\begin{align}
a(\tau)&\simeq& -\frac{1}{H\tau}\label{background}\\
\epsilon\equiv 1-\frac{{\cal H}^{\prime}}{{\cal H}^2}\ll 1\,&,&\qquad \eta\equiv\epsilon-\frac{\epsilon^{\prime}}{2{\cal H}\epsilon}\ll 1\,,\label{slow-roll-def}
\end{eqnarray}
where $H$ is the Hubble parameter during inflation and prime denotes derivative w.r.t. the conformal time $\tau$.

Equation of motion for scalar perturbations, the gauge-invariant Mukhanov-Sasaki variable $u(\tau,y)$,
\begin{equation}\label{u-mukhanov}
u=-z \left( \frac{a^{\prime}}{a}\frac{\delta \phi}{\phi^{\prime}}+\Psi\right), \quad z\equiv \frac{a \phi^{\prime}}{\cal H}, \quad {\cal H}\equiv \frac{a^{\prime}}{a},
\end{equation}
is
\begin{equation}\label{u-eq}
u^{\prime\prime}_k+\left(k^2-\frac{z^{\prime\prime}}{z}\right)u_k=0\,,
\end{equation}
$u_k(\tau)$ is the Fourier mode of $u(\tau,y)$. The most generic solution to \eqref{u-eq} in the leading order in slow-roll parameters $\epsilon, \eta$ may be expressed as:
\begin{equation}\label{u-sol-ds}
u_k(\eta)\simeq\frac{\sqrt{\pi|\tau|}}{2}\left[\AS~  H_{3/2}^{(1)}(k|\tau|)+\BS H_{3/2}^{(2)}(k|\tau|)\right]\,,
\end{equation}
where $H_{3/2}^{(1)}$ and $H_{3/2}^{(2)}$ are respectively  Hankel functions of the first and second kind. The coefficients $\AS$ and $\BS$ are in general scale-dependent and may have non-trivial scale-dependent phases.  They respectively behave like the positive and negative frequency modes. These Bogoliubov coefficients satisfy the normalization condition
\begin{equation}\label{Wronskian}
|\AS(k)|^2-|\BS(k)|^2=1.
\end{equation}
The standard BD vacuum corresponds to $\A=1$ and $\B=0$. However, in general new physics at the onset of inflation can provide us with generic non-BD initial state parameterized with generic $\AS$ and $\BS$. The power spectrum of curvature perturbations is
\begin{equation}
{\mathcal P}_{S}=\frac{k^{3}}{2\pi ^{2}}\left| \frac{u_{k}}{z}\right|^2_{{k/{\cal H}\rightarrow 0}}.
\label{scrpower}
\end{equation}
which for simple chaotic slow-roll models reduce to
\begin{equation}\label{power-spectrum-scalar}
{\mathcal P}_S={\mathcal P}_{BD}\,\gamma_{{}_S} ,
\ee
where
\be
{\mathcal P}_{BD}=\frac{1}{8\pi^2\epsilon}\left(\frac{H}{\mpl}\right)^2,\qquad \gamma_{{}_S}=|\AS-\BS|^2_{{}_{k={\cal H}}}.
\end{equation}

Similarly, one may consider the tensor mode perturbations in a non-BD initial state parameterized by $\AT$ and $\BT$ subject to the normalization condition
\be\label{normalization-Tensor}
|\AT|^2-|\BT|^2=1.
\ee
The power spectrum of tensor modes is then given by \cite{ADSS}
\begin{equation}\label{pwr-tensor}
 {\mathcal P}_{T}={\mathcal P}_{BD}^T\ \gamma_{{}_T}\,,
\ee
where
\be
{\mathcal P}_{BD}^T=\frac{2}{\pi^2}\left(\frac{H}{\mpl}\right)^2\,,\quad \gamma_{{}_T}=|\AT-\BT|^2_{{}_{k={\cal H}}}\,.
\end{equation}

\bibliographystyle{apsrev}

\end{document}